\documentclass{article}
\usepackage{spconf,amsmath,graphicx}
 \usepackage{amsfonts}
 \usepackage{subfig}
 \usepackage{color}
\title{Determined blind source separation via modeling adjacent frequency band correlations in speech signals }
\name{Jianyu Wang$^{1}$, Shanzheng Guan$^{1}$, Zhengqiao Zhao$^{1}$, Nicolas Dobigeon$^{2}$, Jingdong Chen$^{1}$\thanks{Part of this work has been funded by the Artificial Natural Intelligence Toulouse Institute (ANITI, ANR-19-PI3A-0004).}}
\address{$^1$:  CIAIC and Shaanxi Provincial Key Laboratory of Artificial Intelligence, \\
Northwestern Polytechnical University, Xi’an, Shaanxi, China \phantom{Intelli}\\
$^2$: University of Toulouse, IRIT, INP-ENSEEIHT, Toulouse, France}

\begin{document}
\ninept
\maketitle
\begin{abstract}
Multichannel blind source separation (MBSS), which focuses on separating signals of interest from mixed observations, has been extensively studied in acoustic and speech processing. Existing MBSS algorithms, such as independent low-rank matrix analysis (ILRMA) and multichannel nonnegative matrix factorization (MNMF), utilize the low-rank structure of source models but assume that frequency bins are independent. In contrast, independent vector analysis (IVA) does not rely on a low-rank source model but rather captures frequency dependencies based on a uniform correlation assumption. In this work, we demonstrate that dependencies between adjacent frequency bins are significantly stronger than those between bins that are farther apart in typical speech signals. To address this, we introduce a weighted Sinkhorn divergence-based ILRMA (wsILRMA) that simultaneously captures these inter-frequency dependencies and models joint probability distributions. Our approach incorporates an inter-frequency correlation constraint, leading to improved source separation performance compared to existing methods, as evidenced by higher Signal-to-Distortion Ratios (SDRs) and Source-to-Interference Ratios (SIRs).
\end{abstract}
\section{Introduction}
\label{sec:intro}
Multichannel blind source separation (MBSS) involves extracting independent source signals from multichannel observations, where neither the source signals and their statistics nor the mixing process are known in advance \cite{1,2,3}. They can be used in a wide range of acoustic applications including teleconferencing and human-machine speech interfaces. Independent low-rank matrix analysis (ILRMA) is a prominent method for the determined MBSS \cite{9} where the number of sensors exceeds the number of sources. This approach, based on non-negative matrix factorization (NMF), seeks to estimate the demixing matrix by approximating source spectrograms with low-rank matrices. Another NMF-based method, multichannel non-negative matrix factorization (MNMF), models spatial mixing using spatial covariance matrices \cite{10,11}. To enhance computational efficiency, techniques such as FastMNMF \cite{12} and fast full-rank spatial covariance analysis (FastFCA) \cite{13} have been developed. Improvements in separation performance have also been achieved by employing non-Gaussian source models, like Generalized Gaussian and Student-t distributions \cite{14,15,16,17,18,19,20}.Although these methods have achieved some success in estimating the demixing matrix in the STFT domain, they commonly assume that spectral components across different STFT bins (bands) are independent. This assumption often does not hold in practical applications.

Independent vector analysis (IVA) \cite{4,5}, an extension of independent component analysis (ICA) \cite{6}, models statistical dependencies across frequency bins of separated signals. This approach is particularly effective for separating speech signals, where spectral components across different STFT bins often exhibit correlated structures. Additionally, methods that utilize sparse probabilistic priors \cite{21,22}, such as those employing dictionary learning and activation matrices, further enhance separation by leveraging the inherent sparsity of source signals, especially in the STFT domain. However, IVA's reliance on simple statistical dependencies between frequency bins limits its ability to capture more complex relationships, particularly in non-stationary or highly correlated signals like speech \cite{7,8}. How to fully explore spectral dependencies within source models in MBSS remains a challenging issue.  

To address this issue, this work presents a method for refining the source model within the MBSS framework. We introduce a novel approach, called weighted Sinkhorn-based ILRMA (wsILRMA), which utilizes NMF for source modeling while employing Sinkhorn divergence \cite{27,28,29} to model the inter-frequency dependencies of the squared magnitude spectra \cite{30}. This method relaxes the conventional assumption of frequency independence inherent in the standard ILRMA. Unlike previous Sinkhorn divergence-based source models \cite{34}, our approach more effectively captures non-linear spectral structures and aligns with the joint time-frequency representation of signals. Specifically, it incorporates a regularization term that accounts for time-frequency coherence \cite{31}, constraining the transport matrix to improve the modeling of spectral dependencies. This enhancement results in greater source model accuracy and better separation performance, as demonstrated by numerical results from simulations.

\section{Signal Model and Problem Formulation}
\label{sec:format}

We consider a determined MBSS problem involving $N$ sources and $M$ microphones. For simplicity, we assume $N=M$, though the method developed here can be extended to the more general case where $N\leq M$. The convolutive mixture in the time domain can be reformulated into the STFT domain as follows:
\begin{align}
\mathbf{x}(f,t) = & \mathbf{A}(f) \mathbf{s}(f,t),
\end{align}
with
\begin{align}
\mathbf{x}(f,t) &= \left[ \begin{array}{cccc}
                           x_1(f,t) & x_2(f,t) & \cdots & x_N(f,t)
                         \end{array} \right]^\mathsf{T}, \\
\mathbf{s}(f,t) &= \left[ \begin{array}{cccc}
                           s_1(f,t) & s_2(f,t) & \cdots & s_N(f,t)
                         \end{array} \right]^\mathsf{T},
\end{align}

The primary challenge in the separation process is accurately estimating the demixing matrix $\mathbf{W}(f)$ to recover the source signals, i.e.,
\begin{align}\label{demixing}
\mathbf{y}(f,t) = \mathbf{W}(f) \mathbf{x}(f,t),
\end{align}
where $\mathbf{y}(f,t)$ denotes an estimate of  $\mathbf{s}(f,t)$ with
\begin{align}
\mathbf{y}(f,t) &= \left[ \begin{array}{cccc}
                           y_1(f,t) & y_2(f,t) & \cdots & y_N(f,t)
                         \end{array} \right]^\mathsf{T}, \\
\mathbf{W}(f) &= \left[ \begin{array}{cccc}
                           \mathbf{w}_1(f) & \mathbf{w}_2(f) & \cdots & \mathbf{w}_N(f)
                         \end{array} \right], \\
\mathbf{w}_n(f) &= \left[ \begin{array}{cccc}
                           w_{n,1}(f) & w_{n,2}(f) & \cdots & w_{n,N}(f)
                         \end{array} \right]^\mathsf{T},
\end{align}
and $n$ denoting the $n$th source at time $t$.

In MBSS, the statistical distribution of source signals is vital for algorithm design and performance. Most BSS algorithms operate under the assumption that the sources are non-Gaussian, while the mixed signals often approximate a Gaussian distribution due to the central limit theorem. Modeling non-Gaussian sources directly can be challenging; however, the Boltzmann distribution provides a practical approach, enabling the modeling of sources as a multivariate distribution \cite{32}, i.e., 
\begin{align}
p\left[ \mathbf{s}_n(:,t) \right] \propto \exp\left[{-G(\mathbf{s}_n(:,t))}\right]
\end{align}
where $G(\cdot)$ denotes a contrast function \cite{33}.

Another fundamental aspect of MBSS is the statistical independence of source signals, which is crucial for estimating the demixing matrix in blind source separation algorithms. Typically, Typically, MBSS algorithms achieves source independence by minimizing the mutual information (KL divergence) between the demixed signals: 
\begin{align}
\label{kl}
\mkern-10mu \mathcal{L} & = \mathcal{KL}\left( p[\mathbf{y}_1(:,t),\cdots,\mathbf{y}_N(:,t)] \bigg| \prod_{n=1}^{N} p\left[ \mathbf{y}_n(:,t) \right] \right) \nonumber \\
& = \!\! \int \!\! p\!\left[\mathbf{y}_{\!1\!}(:,\!t),\!\cdots\!,\!\mathbf{y}_{\!N\!}(:,t)\right] \!\log \! \frac{p\!\left[\mathbf{y}_{\!1\!}(:,\!t),\!\cdots\!,\!\mathbf{y}_{\!N\!}(:,\!t)\right]}{\prod_{n=1}^{N} p\left[ \mathbf{y}_n(:,t) \right]} d\mathbf{y}_{\!1\!}\! \cdots \!d\mathbf{y}_{\!N\!} \nonumber \\
& = \sum_{n=1}^N \mathcal{H}\left[ \mathbf{y}_n(:,t) \right] - \mathcal{H}\left[ \mathbf{y}_1(:,t),\cdots,\mathbf{y}_N(:,t) \right],
\end{align}
where $\mathcal{H}\left[\cdot\right]$ denotes the entropy.
Using \eqref{demixing}, the entropy related to the joint distribution can be rewritten as
\begin{align}
& \mkern-10mu  \mathcal{H}\left[ \mathbf{y}_1(:,t),\cdots,\mathbf{y}_N(:,t) \right] = - \int  p  \left[\mathbf{x}_{1}(:,t),\cdots,\mathbf{x}_{M}(:,t)\right] \nonumber \\
& \mkern-10mu  \times \log p\left[\mathbf{x}_{1}(:,t),\cdots,\mathbf{x}_{M}(:,t)\right] d\mathbf{x}_{1}(:,t) \cdots d\mathbf{x}_{M}(:,t)  \nonumber \\
& +  \sum_{f=1}^F  \log\det\mathbf{W}(f), \!\!\!\!
\end{align}
\vspace{-0.1cm}
Finally, the cost function in (\ref{kl}) can be expressed as
\begin{align}\label{ObjectFunc}
 \mathcal{L} = \mathit{const} - \sum_{f=1}^F \log\det\mathbf{W}(f) - \sum_{n=1}^N G(\mathbf{y}_n(:,t)),
\end{align}
where $G\left[\mathbf{y}_n(:,t)\right] = \mathbb{E}\left\{ \log p\left[\mathbf{y}(:,t)\right] \right\}$ is the contrast function associated
 with the estimated separated source signals.

\section{Proposed Model and Source Separation algorithm}
In this section, we begin by examining the inter-band correlation in the STFT domain. Next, we describe the Sinkhorn divergence-based contrast function, and conclude by introducing a constant constraint to this contrast function to better capture non-linear inter-band dependencies.

\subsection{Illustration of inter-band correlation}
\begin{figure}[t!]
\includegraphics[width=8cm]{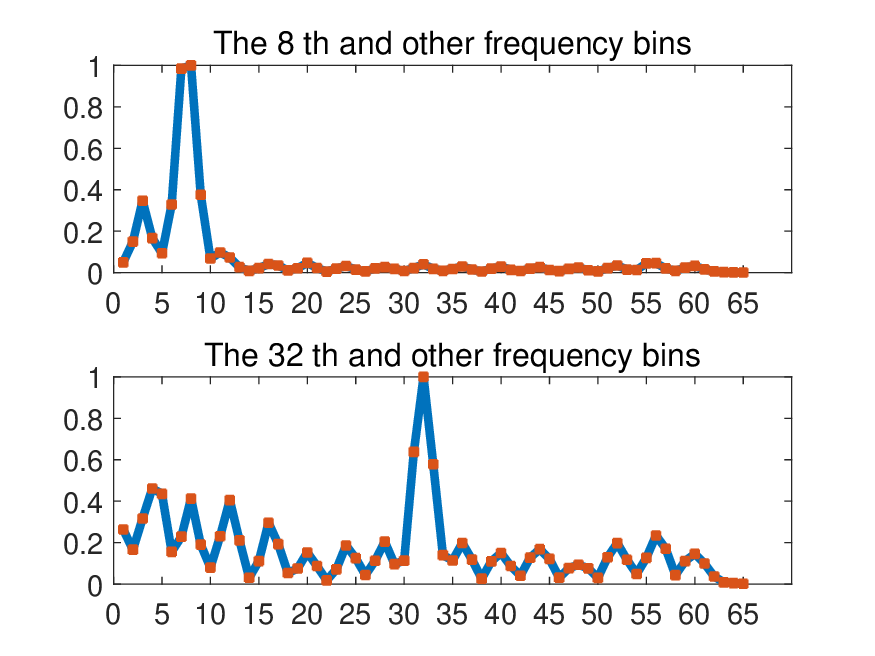}
\caption{The magnitude of pairwise normalized correlation coefficients between STFT frequency bins of a speech signal. The sampling rate is $8$ kHz, the frame length is
$16$ ms ($128$ points), the FFT length is $128$, and the overlap is $50\%$}
\label{fig1}
\end{figure}

For stationary signals with a sufficiently long STFT length, spectral components from different STFT bins are generally expected to be uncorrelated. However, in MBSS scenarios involving speech signals and time-varying acoustic environments, the STFT length is often limited and neighboring frames frequently overlap. Consequently, correlations can arise between different frequency bins, especially among neighboring bins. To illustrate this, we plot in \ref{fig1} the magnitude of pairwise normalized correlation coefficients between STFT frequency bins of a speech signal, where the normalized correlation coefficient between two frequencies bins is defined as 
\begin{align}
{r}_x(f_1,f_2) = \frac{\mathbb{E}\left[ x(f_1,t) x^{\mathsf{H}}(f_2,t) \right]}{\sqrt{\mathbb{E}\left[|x(f_1,t)|^2\right]} \sqrt{\mathbb{E}\left[|x(f_2,t)|^2\right]}},
\end{align}
where $f_1$ and $f_2$ denotes two frequency bins, and $^{\mathsf{H}}$ is the conjugate transpose operator. 

As shown in Fig.~\ref{fig1}, the spectral components from neighboring STFT bins exhibit strong dependencies. Therefore, it is crucial to account for these dependencies when developing MBSS algorithms, as they can significantly affect performance.

\subsection{Contrast function based on Optimal spectral transport}
Unlike traditional contrast functions in BSS, such as those used in ICA that rely on non-Gaussianity, Sinkhorn divergence $\mathcal{S}(\cdot|\cdot)$ can manage a broad spectrum of complex and multimodal distributions, providing greater flexibility. This work leverages this flexibility by using Sinkhorn divergence to optimally project the spectral components of each source, thereby developing a contrast function for source reconstruction. Specifically, we consider:
\begin{align}\label{Wasserstein}
\!\!\mathcal{S}_{\!\frac{1}{\lambda}\!\!}\left[ \tilde{\mathbf{y}}_{\!n\!}(:,\!t), \tilde{\boldsymbol{\sigma}}_{\!n\!}(:,\!t) \! \right] \!\! = \!\!\! \min_{\!\mathbf{Q}\in\Pi(|\tilde{\mathbf{y}}_n(:,\!t)|^2,\tilde{\boldsymbol{\sigma}}^2_n(:,\!t))} \! \left[ \! \langle \mathbf{Q}, \mathbf{C} \rangle \! - \! \frac{1}{\lambda} \mathcal{H}(\mathbf{Q}) \! \right],
\end{align}
where the transport path $\Pi(\cdot,\cdot)$, the cost matrix $\mathbf{C}$, the normalized two variables $|\tilde{y}_n(f_1,t)|^2$ and $\tilde{\boldsymbol{\sigma}}^2$ are defined respectively as
\begin{align}
\!\!\! \Pi\left(\!|\tilde{\mathbf{y}}_{\!n\!}(:,\!t)|^{\!2}\!\!,\!\tilde{\boldsymbol{\sigma}}^2_{\!n\!}(:,\!t)\!\right) \!\! & = \!\! \left\{ \mathbf{Q}\!\in\!\!\mathbb{R}_+^{\!F\!\times F\!}\!\!\!: \mathbf{Q}\mathbf{1} \! = \! |\tilde{\mathbf{y}}_{\!n\!}(:,\!t)|^{\!2}\!\!, \mathbf{Q}^\mathsf{T} \! \mathbf{1} \! = \! \tilde{\boldsymbol{\sigma}}^2_{\!n\!}(:,\!t)\! \right\}, \nonumber\\
\left[\mathbf{C}\right]_{f_1,f_2} & = \left( \log\frac{|\tilde{y}_n(f_1,t)|^2}{\tilde{\sigma}^2_n(f_2,t)} \right)^2, \nonumber\\
|\tilde{y}_n(f_1,t)|^2 & = \frac{|y_n(f_1,t)|^2}{\sum_{f_1=1}^F|y_n(f_1,t)|^2}, \nonumber \\
\tilde{\sigma}_n^2(f_2,t) & = \frac{\sigma_n^2(f_2,t)}{\sum_{f_2=1}^F\sigma_n^2(f_2,t)}. \nonumber
\end{align}

\subsection{The weighted Sinkhorn divergence-based ILRMA (wsILRMA)}
\begin{figure}[!t]
  \centering
  \subfloat[$\mathbf{U}$]{\includegraphics[width=1.6in]{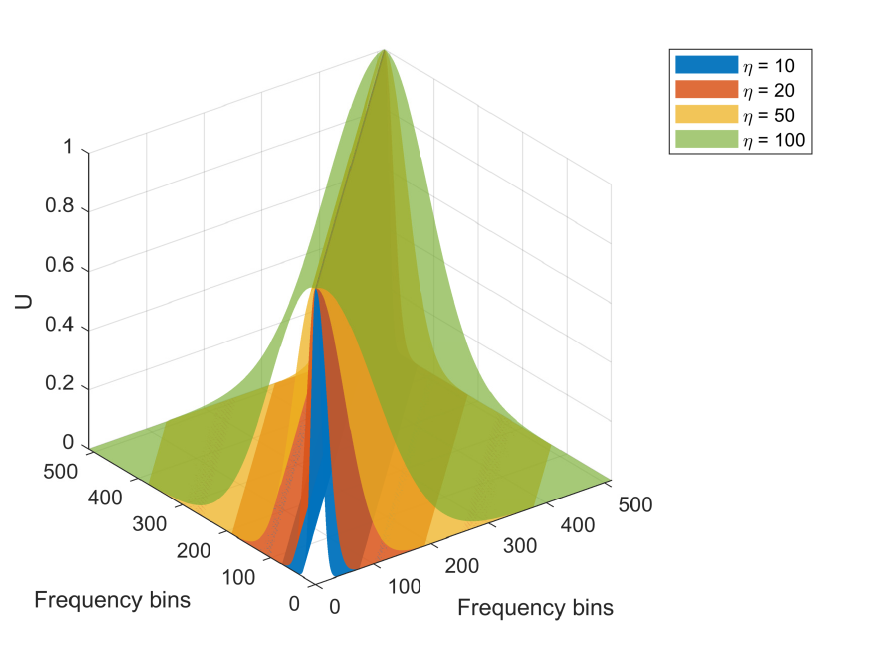}%
  \label{Reuters_delta_Coh}}
  \hfil
  \subfloat[$-\log \mathbf{U}$]{\includegraphics[width=1.6in]{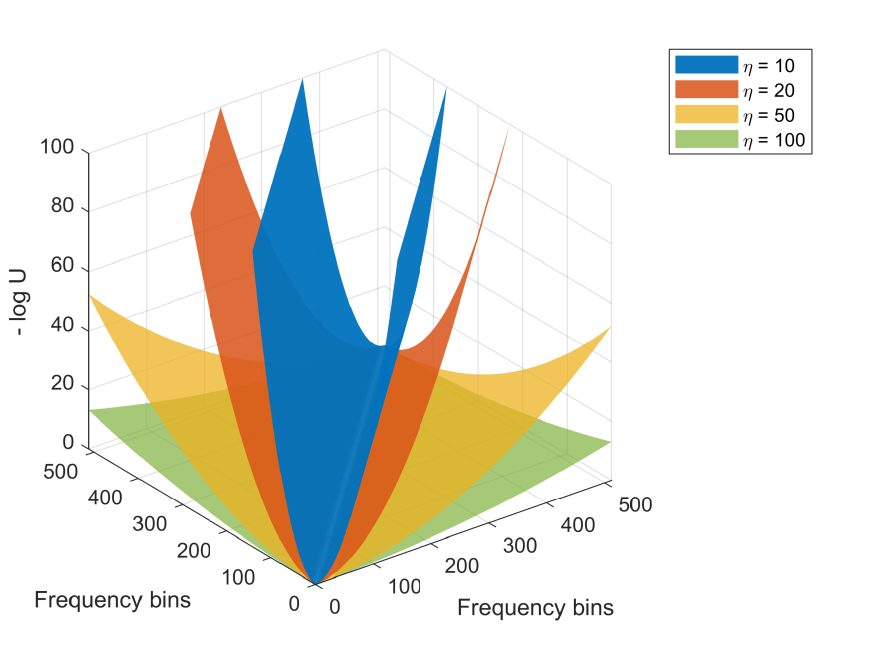}%
  \label{Reuters_delta_SC}}
  \hfil
\caption{Visualization of the fixed amplitude weights $\mathbf{U}$ (a) and $-\log\mathbf{U}$ (b). The FFT length is $1024$ points. $\mathbf{U}$ reflect similar inter-band dependencies as the inter-band correlation coefficients.}
\label{fig2}
\end{figure}

By employing the Sinkhorn divergence-based contrast function, the final term in \eqref{ObjectFunc} can be formulated as estimating the optimal mapping from source distribution to the reconstructed signals $\mathbf{y}_n(:,t)$:
\begin{align}
\hat{\boldsymbol{\sigma}}^2_n(:,t) = \left(\hat{\mathbf{Q}}_{n,t}^\mathsf{T} \mathbf{1}\right) \cdot \sum_{f=1}^F |y_n(f,t)|^2,
\end{align}
where
\begin{align}
\hat{\mathbf{Q}}_{n,t} = \arg\min \mathcal{S}_{\frac{1}{\lambda},\gamma}\left( |\tilde{\mathbf{y}}_n(:,t)|^2, \tilde{\boldsymbol{\sigma}}_n^2(:,t) \right),
\end{align}
In the above definition, the function $\mathcal{S}_{\frac{1}{\lambda},\gamma}\left(\cdot \big| \cdot\right)$ denotes the Sinkhorn divergence as defined in \eqref{Wasserstein}. We further introduce a fixed amplitude weights to capture the inter-band dependencies illustrated in Fig.~\ref{fig1} and define the proposed weighted Sinkhorn divergence-based objective function as:  
\begin{align}\label{SinkhornDiv}
& \mathcal{S}_{\!\frac{1}{\lambda}\!,\gamma\!\!}\left( |\tilde{\mathbf{y}}(\!:,\!t)|^2\!, \tilde{\boldsymbol{\sigma}}^2_{\!n}(\!:,\!t)\! \right) \! = \! \left[ \left\langle \mathbf{Q}_{n\!,t}, \mathbf{C}_{\!n,t} - \log\mathbf{U}  \right\rangle \! - \! \frac{1}{\lambda}\mathcal{H}[\mathbf{Q}_{n\!,t}] \right] \nonumber \\
& + \gamma \left[ \mathcal{L}_\phi \left(  \mathbf{Q}_{n\!,t}^\mathsf{T}\mathbf{1} \Big| \tilde{\boldsymbol{\sigma}}^2_{\!n}(:,\!t) \! \right) + \mathcal{L}_\phi \left( \mathbf{Q}_{n\!,t} \mathbf{1} \Big| |\mathbf{y}_{\!n}(:,\!t)|^2 \right) \right],
\end{align}
where $\mathcal{L}_{\phi}(\cdot\big|\cdot)$ denotes a distance measure, chosen in this work as the KL divergence, and the term $\mathbf{U}$ stands for a fixed amplitude constant introduced to adjust the cost matrix such that the resulting transport matrix effectively captures the inter-band dependencies, similar to the inter-band correlation illustrated in Fig.~\ref{fig1}, which is defined as
\begin{align}
\left[\mathbf{U}\right]_{f_{\!1},\!f_{\!2}} \!\! = \! \frac{1}{\max(\mathbf{U})}\!\cdot\!\frac{1}{\sqrt{2\pi}\eta} \! \cdot \! \exp\!\left( \! -\frac{\left({|f_1 - f_2|}\right)^2}{2\eta^2} \! \right),
\end{align}
where $\eta$ reflects the width of inter-band frequency dependencies. 
In other words,  $-\log\mathbf{U}$ ensures that the transport matrix accurately reflects the dependencies across different frequency bands, aligning with the desired spectral properties. Figure~\ref{fig2} shows the resulting shapes of $\mathbf{U}$ and $-\log\mathbf{U}$ with respect to different choices of $\eta$.

Deriving the gradient of \eqref{SinkhornDiv} and setting it to zero yields the solution
\begin{align}
\hat{\mathbf{Q}}_{n,t} = \mbox{diag}\left( \boldsymbol{\nu}_{n,t} \right) \mathbf{K}_{n,t}\mbox{diag}\left( \boldsymbol{\xi}_{n,t} \right),
\end{align}
where
\begin{align}
 & \mathbf{K}_{n,t} = \mathbf{U}^{\lambda} \cdot \exp\left( -\lambda \mathbf{C}_{n,t} - 1 \right), \\
 & \boldsymbol{\nu}_{n,t} = \left[ \frac{|\mathbf{y}_n(:,t)|^2}{\mathbf{K}_{n,t}^\mathsf{T} \boldsymbol{\xi}_{n,t}} \right]^{\frac{\lambda\gamma}{\lambda\gamma + 1}}, \\
 & \boldsymbol{\xi}_{n,t} = \left[ \frac{\boldsymbol{\sigma}_{n,t}^2}{\mathbf{K}_{n,t} \left[ \frac{|\mathbf{y}_n(:,t)|^2}{\mathbf{K}_{n,t}^\mathsf{T} \boldsymbol{\xi}_{n,t}} \right]^{\frac{\lambda\gamma}{\lambda\gamma + 1}}} \right]^{\frac{\lambda\gamma}{\lambda\gamma + 1}}.
\end{align}
Given that the components of $\boldsymbol{\sigma}^2$ in the optimal spectral transport contrast function are non-negative, NMF is especially suitable for modeling them. We decompose the components of $\boldsymbol{\sigma}^2$ as
\begin{equation}
    \sigma_{n,f,t}^2 = \sum_{k=1}^K u_{n,f,k} v_{n,k,t}.
\end{equation}
Using the decomposition, we can derive the following parameter update rules:
\begin{align}
 & u_{n,f,k} \leftarrow \sqrt{\frac{\sum_t \left[ \hat{\mathbf{Q}}_{n,t} \mathbf{1} \right]_f v_{n,k,t} \left( \sum_{k^\prime} u_{n,f,k^\prime} v_{n,k^\prime,t}\right)^{-2} }{\sum_t \left[ \hat{\mathbf{Q}}_{n,t} \mathbf{1} \right]_f \left( \sum_{k^\prime} u_{n,f,k^\prime} v_{n,k^\prime,t}\right)^{-1}}}, \\
 & v_{n,k,t} \leftarrow \sqrt{\frac{\sum_f \left[ \hat{\mathbf{Q}}_{n,t} \mathbf{1} \right]_f u_{n,f,k} \left( \sum_{k^\prime} u_{n,f,k^\prime} v_{n,k^\prime,t}\right)^{-2}}{\sum_f \left[ \hat{\mathbf{Q}}_{n,t} \mathbf{1} \right]_f \left( \sum_{k^\prime} u_{n,f,k^\prime} v_{n,k^\prime,t}\right)^{-1}}},
\end{align}

Finally, the demixing matrix $\mathbf{W}(f)$ of wsILRMA is iteratively updated following the same strategy adpoted by IVA, i.e.,
\begin{align}
& \mathbf{O}_{n,f} = \frac{1}{T}\sum_t \frac{1}{\sum_k u_{n,f,k}v_{n,k,t}} \mathbf{x}\left(f,t\right) \mathbf{x}^{\mathsf{H}}\left(f,t\right), \\
& \mathbf{w}_n(f) \leftarrow \left[ \mathbf{W}(f) \mathbf{O}_{n,f} \right]^{-1} \mathbf{e}_n, \\
& \mathbf{w}_n(f) \leftarrow \mathbf{w}_n(f) \left[ \mathbf{w}_n^\mathsf{H}(f) \mathbf{O}_{n,f} \mathbf{w}_n(f) \right]^{-\frac{1}{2}},
\end{align}
where $\mathbf{e}_n$ denotes the $n$th column of the identity matrix.

\begin{figure}[t]
\begin{minipage}[b]{1\linewidth}
  \centering
  \centerline{\includegraphics[width=9cm]{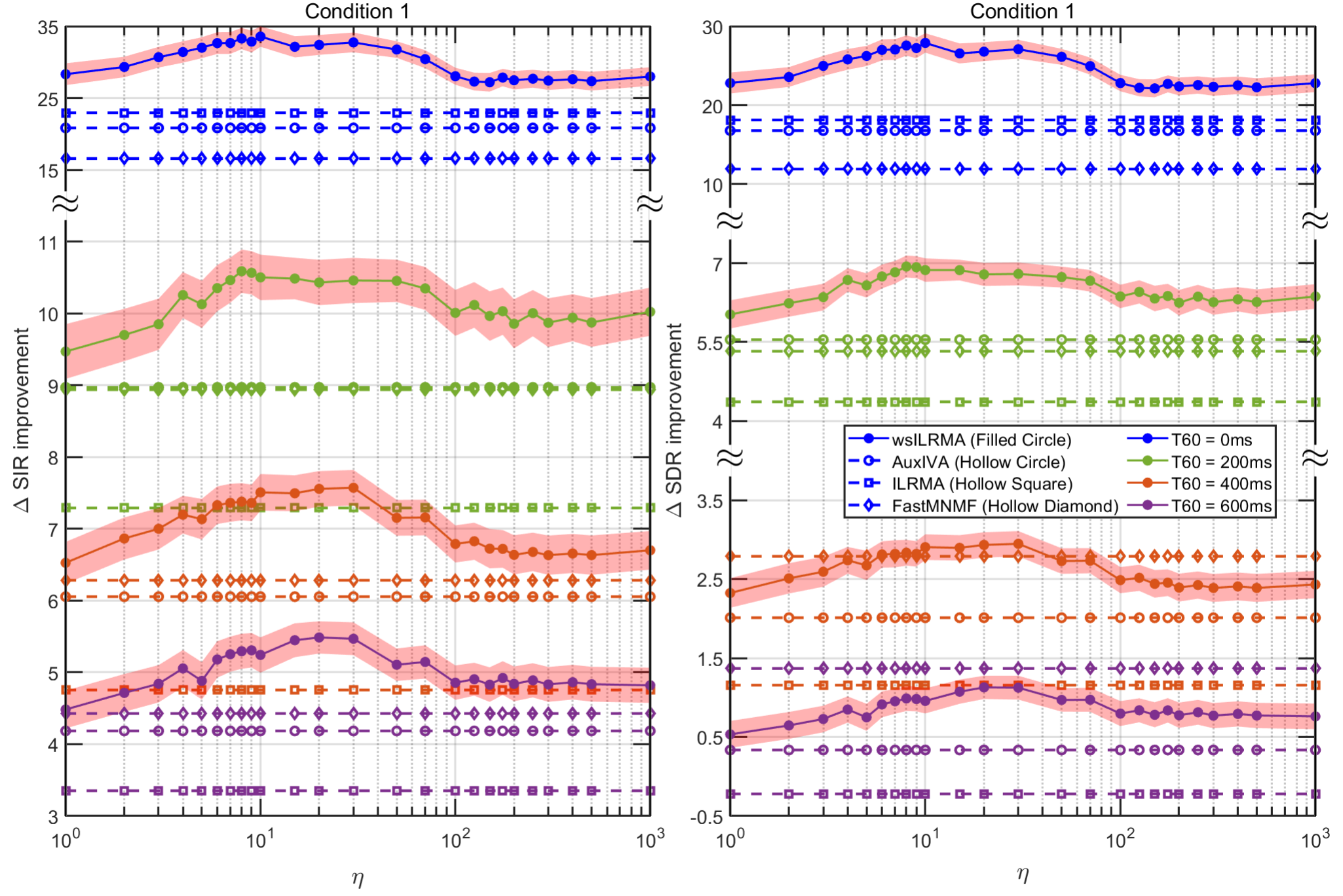}}
\end{minipage}
\caption{Simulation results for MBSS in \textbf{Condition 1}. Average SIR (left) and SDR (right) performance with varying amplitude weighting parameter $\eta$ under different reverberation conditions. Note that the x-axis in the logarithmic scale. The bands show the $95\%$ confidence interval around the mean. The dashed lines indicate the mean performance for comparison methods.}
\label{fig3}
\end{figure}

\begin{figure}[t]
\begin{minipage}[b]{1\linewidth}
  \centering
  \centerline{\includegraphics[width=8.97cm]{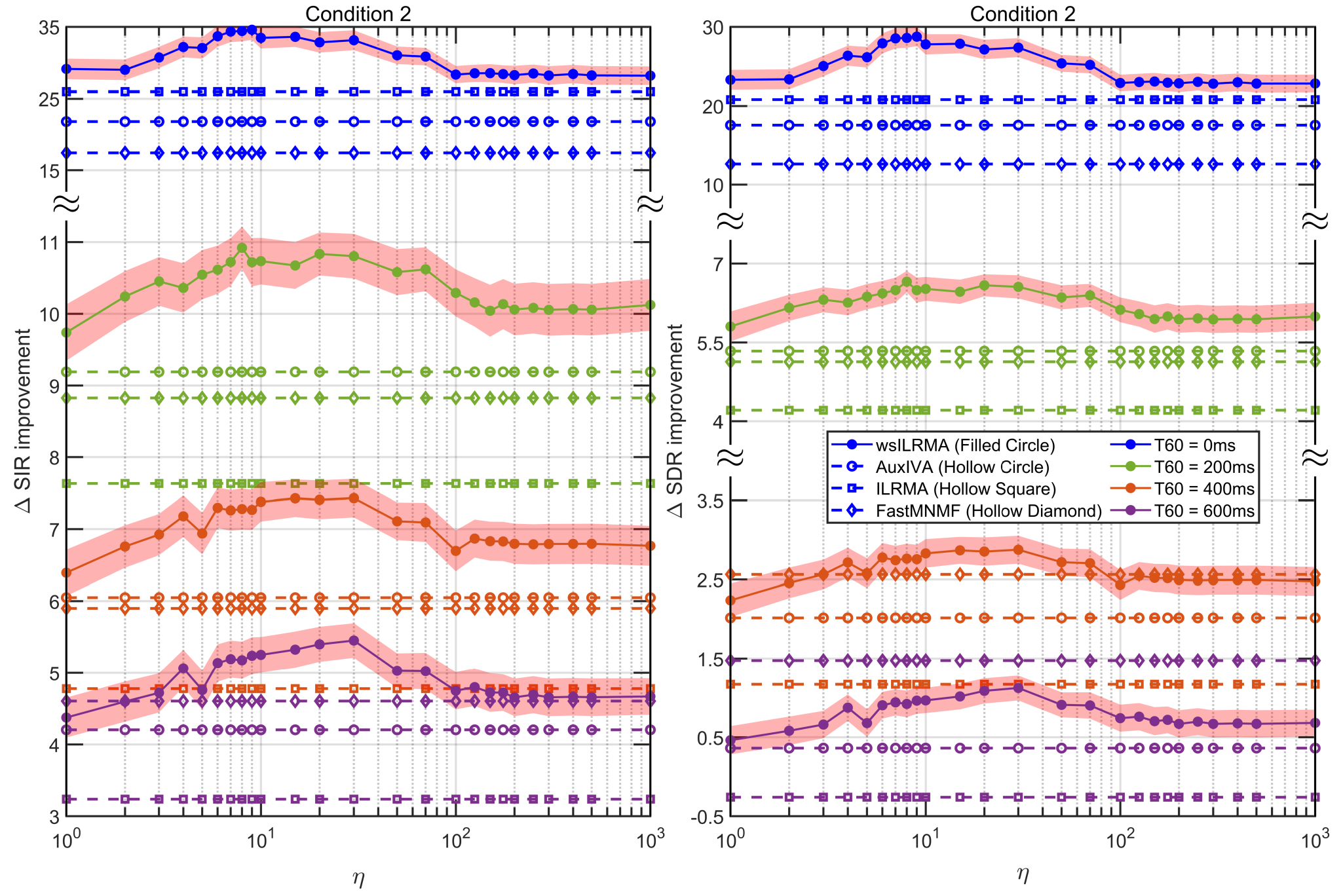}}
\end{minipage}
\caption{Simulation results for MBSS in \textbf{Condition 2}. Average SIR (left) and SDR (right) performance with varying amplitude weighting parameter $\eta$ under different reverberation conditions. Note that the x-axis in the logarithmic scale. The bands show the $95\%$ confidence interval around the mean. The dashed lines indicate the mean performance for comparison methods.}
\label{fig4}
\end{figure}

\section{Simulation results}
\label{sec:pagestyle}
This section describes the simulation configure, and then the simulation results and discussions.

\subsection{Simulation configuration}

To simulate the MBSS experimental data, we use clean speech signals from the Wall Street Journal (WSJ0) database and follow the SISEC challenge configuration to generate mixed signals. We set the number of sources and microphones to 2, i.e., $M = N = 2$. The simulated room measures $8\times 8\times 3$ meters, with two microphones placed at the center, spaced $6$~cm apart. Two sets of source positions are used, creating 2 evaluation conditions. \textbf{Condition 1}, the sound sources are positioned 2 meters from the microphones at angles of 10° and 20°, respectively. \textbf{Condition 2}, the sources are still 2 meters away but at wider angles of 45° and 55°.

The room impulse responses are generated using the image source model, with sound absorption coefficients calculated based on Sabine’s formula. The reverberation time 
$T_{60}$ is set to values of $\{0, 200, 400, 600\}$ ms. For each configuration combination (three or four, depending on the simulation) and each 
$T_{60}$ value, 100 mixtures are generated to assess separation performance. The sampling rate is set to $16$~kHz.
  
\subsection{Algorithm parameters} 

During the experiments, the hyperparameters of wsILRMA for all simulations are set to: $\lambda = 4$, $\gamma = 1$, and $K = 10$. The hyperparameter $\eta$ is selected from \{1, 2, 3, 4, 5, 6, 7, 8, 9, 10, 15, 20, 30, 50, 70, 100, 125, 150, 175, 200, 250, 300, 400, 500, 1000\} for testing. 

\subsection{Compared algorithms and performance metrics  } 
The following widely used competing algorithms are also evaluated for comparison: AuxIVA \cite{5}, ILRMA \cite{9}, and FastMNMF \cite{12}. Signal-to-distortion ratio (SDR) and source-to-interference ratio (SIR)  are adopted as the performance metrics \cite{35}.

\subsection{Simulation results and discussions}
It is evident that wsILRMA consistently outperforms other algorithms in terms of SIR and SDR across both experimental conditions, as illustrated in Figs.~\ref{fig3} and \ref{fig4}, particularly at lower reverberation times (0 ms and 200 ms). The results also highlight that the performance of wsILRMA is sensitive to the choice of 
$\eta$, with the optimal range being approximately 10 to 100. This suggests that accounting for dependencies between adjacent frequency bands enhances separation performance.

Furthermore, as reverberation time increases, the performance of all MBSS algorithms deteriorates. Nonetheless, wsILRMA remains one of the most robust algorithms even in environments with high reverberation. Overall, by incorporating inter-band dependencies, wsILRMA delivers the best performance under complex experimental conditions.

\section{Conclusion}
In acoustic and speech applications, MBSS is commonly performed in the STFT domain to efficiently handle convolutive mixing. In this domain, spectral components from different frequency bins can exhibit significant dependencies, which are often overlooked by existing MBSS algorithms. To address this issue, this paper presented a weighted Sinkhorn-based ILRMA (wsILRMA). By utilizing Sinkhorn divergence to capture non-linear inter-frequency dependencies, wsILRMA overcomes the limitations of frequency independence in current methods, resulting in improved separation performance. Future work will aim to extend the model to handle more complex environments.

\vfill\pagebreak

\bibliographystyle{IEEEbib}
\bibliography{strings,refs}

\end{document}